\pdfoutput=1

\documentclass[aip,jcp,reprint,footinbib,citeautoscript]{revtex4-1}
\usepackage{graphicx} 
\usepackage{amsmath}
\usepackage[version=3]{mhchem} 
\usepackage{mathtools} 
\usepackage{bm}
\usepackage{amssymb} 
\usepackage{tabularx,multirow}
\usepackage{siunitx}

\usepackage{hyperref}
\hypersetup{colorlinks=true,linkcolor=blue,citecolor=blue,urlcolor=blue}

\newcommand{\secref}[1]{Sec.~\ref{sec:#1}}

\newcommand{\pder}[3][]{\frac{\partial^{#1}{#2}}{\partial{#3}^{#1}}}
\newcommand{\pders}[3]{\frac{\partial^2{#1}}{\partial{#2}\partial{#3}}}
\newcommand{\eqn}[1]{Eq.~\ref{#1}}

\newcommand{\Ref}[1]{Ref.~\onlinecite{#1}}
\newcommand{\rmd}{\mathrm{d}}

\begin{document}

\title{\textit{Ab initio} instanton rate theory %
made efficient using Gaussian process regression}

\author{Gabriel Laude}
\affiliation{Laboratory of Physical Chemistry, ETH Zurich, 8093 Zurich, Switzerland}
\affiliation{On exchange from School of Chemistry, University of Edinburgh, UK}
\author{Danilo Calderini}
\affiliation{Laboratory of Physical Chemistry, ETH Zurich, 8093 Zurich, Switzerland}
\author{David P. Tew} 
\affiliation{Max-Planck-Institut f\"ur Festk\"orperforschung, Heisenbergstra{\ss}e 1, 70569 Stuttgart, Germany}
\author{Jeremy O. Richardson}
\email{jeremy.richardson@phys.chem.ethz.ch}
\affiliation{Laboratory of Physical Chemistry, ETH Zurich, 8093 Zurich, Switzerland}

\date{\today}

\begin{abstract}
\textit{Ab initio} instanton rate theory is a computational method
for rigorously including tunnelling effects into calculations
of chemical reaction rates
based on a potential-energy surface computed on the fly from electronic-structure theory.
This approach is necessary to extend conventional transition-state theory into the deep-tunnelling regime,
but is also more computationally expensive as it requires many more \textit{ab initio} calculations.
We propose an approach which uses Gaussian process regression to fit the potential-energy surface locally around the dominant tunnelling pathway.
The method can be converged to give the same result as from an on-the-fly \textit{ab initio} instanton calculation
but requires far fewer electronic-structure calculations.
This makes it a practical approach for obtaining accurate rate constants based on high-level electronic-structure methods.
We show fast convergence to reproduce benchmark \ce{H + CH4} results
and evaluate new low-temperature rates of \ce{H + C2H6} in full dimensionality at a UCCSD(T)-F12b/cc-pVTZ-F12 level.
\end{abstract}

\maketitle

\section{Introduction}
\label{sec:intro}
Transition-state theory (TST) has surely become the most popular method for evaluating reaction rates in gas-phase chemistry.
\cite{Truhlar1996TST}
It has achieved this status due to its simplicity and the fact that it can be evaluated with efficient computational algorithms.
Two geometry optimisations are needed, for the reactant and transition states
and two Hessian calculations, one at each stationary point.
As only a small number of electronic-structure calculations are needed to evaluate the TST rate,
expensive high-level \textit{ab initio} methods can be used.
This is necessary to achieve a good prediction, as small errors in the PES lead to exponential errors in the rate.
TST however is based on classical dynamics and neglects important quantum effects such as tunnelling,
\cite{Wigner1938TST}
which can dominate the mechanism of certain chemical reactions of interest. \cite{BellBook,Carpenter2011tunnelling,Ley2012tunnelling,Meisner2016review}

Ring-polymer instanton theory
has proved itself to be a useful and accurate method for computing the rate
of a chemical reaction dominated by tunnelling.
\cite{Perspective}
The method is based on a first-principles derivation from
the path-integral representation of the quantum rate
\cite{Miller1975semiclassical,AdiabaticGreens,InstReview}
and can be thought of as a quantum-mechanical generalisation of TST\@.
A ring-polymer discretisation of the path integral allows a simple optimisation algorithm to be used
for locating the dominant tunnelling pathway, known as the ``instanton''.
\cite{RPInst,Andersson2009Hmethane,Rommel2011locating,InstReview}
As with TST, it is possible to combine the instanton method with \textit{ab initio} electronic-structure calculations
to evaluate the potential-energy surface (PES) on the fly.
\cite{Milnikov2004,HCH4,Asgeirsson2018instanton,Goumans2010Hbenzene,Kryvohuz2012abinitio}
When compared with benchmark quantum dynamics approaches applied to polyatomic reactions,
the instanton method typically gives low-temperature rates within about $20-30\%$ of an exact calculation on the same PES. \cite{HCH4,MUSTreview}
This is, in many cases, less than the the error in the rate which can be expected to result from the best achievable convergence
of the electronic Schr\"odinger equation,
implying that the accuracy of instanton theory itself is not the major issue.

The \textit{ab initio} instanton method is very efficient when compared with other quantum dynamics approaches,
including path-integral molecular dynamics
or wave-function propagation.
However, it remains considerably more computationally expensive than a TST calculation.
The major reason for this expense is that energies, gradients and Hessians of the PES are required,
not just at the transition state,
but for each ring-polymer bead along the instanton, of which about 100 may be required.
For high-accuracy electronic-structure methods, such as is provided by coupled-cluster theory,
gradients and Hessians are typically evaluated using finite-differences, and can thus consume a lot of computational power.
If the ring-polymer instanton method is to become widely applied in place of TST,
the number of \textit{ab initio} points will need to be reduced to bring the computational expense down,
closer to that of a TST calculation.

It is important that high-quality electronic-structure calculations are employed
as results can be strongly-dependent on the PES
and give significant errors when using cheaper and less-accurate surfaces. \cite{Yagi2004malonaldehyde,MUSTreview}
One suggestion for decreasing the computational effort required is to run the instanton calculation
using a low-level surface and partially correct the result using a few high-level single-point calculations along the optimised pathway.
\cite{Milnikov2003,Milnikov2004,Meisner2018dual}
This approach (termed the `dual-level instanton approach') certainly improves results, but cannot always been relied upon
as, in certain cases, the location of the instanton pathway may vary considerably depending on the quality of the PES.
One can also use Taylor series expansions around the stationary points to obtain an approximate instanton solution analytically.
\cite{Miller1990SCTST,Nguyen2010SCTST,Greene20161D,Greene2016SCTST,Smedarchina2012rainbow}
These approaches also have the potential to break down when the instanton pathway exhibits strong corner-cutting behaviour
and deviates significantly from the transition state.

The procedure which has generally been followed for
ring-polymer molecular dynamics rate theory \cite{RPMDgasPhase,Suleimanov2016rate}
or wave-function propagation methods \cite{Althorpe+Clary2003review,Fu2017scattering}
has been to use an analytical function 
for the PES which is fitted to approximately reproduce \textit{ab initio} points on the surface.
In particular much attention has been given to water potentials
\cite{Wang2010water,Babin2014MBpol,Nguyen2018water}
on which instanton calculations have also been carried out for comparison with high-resolution spectroscopy.
\cite{tunnel,hexamerprism}
Despite improvements and automation of this procedure, 
it remains a difficult task to fit a global potential, and is often based on tens of thousands of \textit{ab initio} points,
\cite{Bowman2015CH3CHOO}
computations which we wish to avoid.

\begin{figure}[t]
	\centering
	\includegraphics[width=\columnwidth, keepaspectratio]{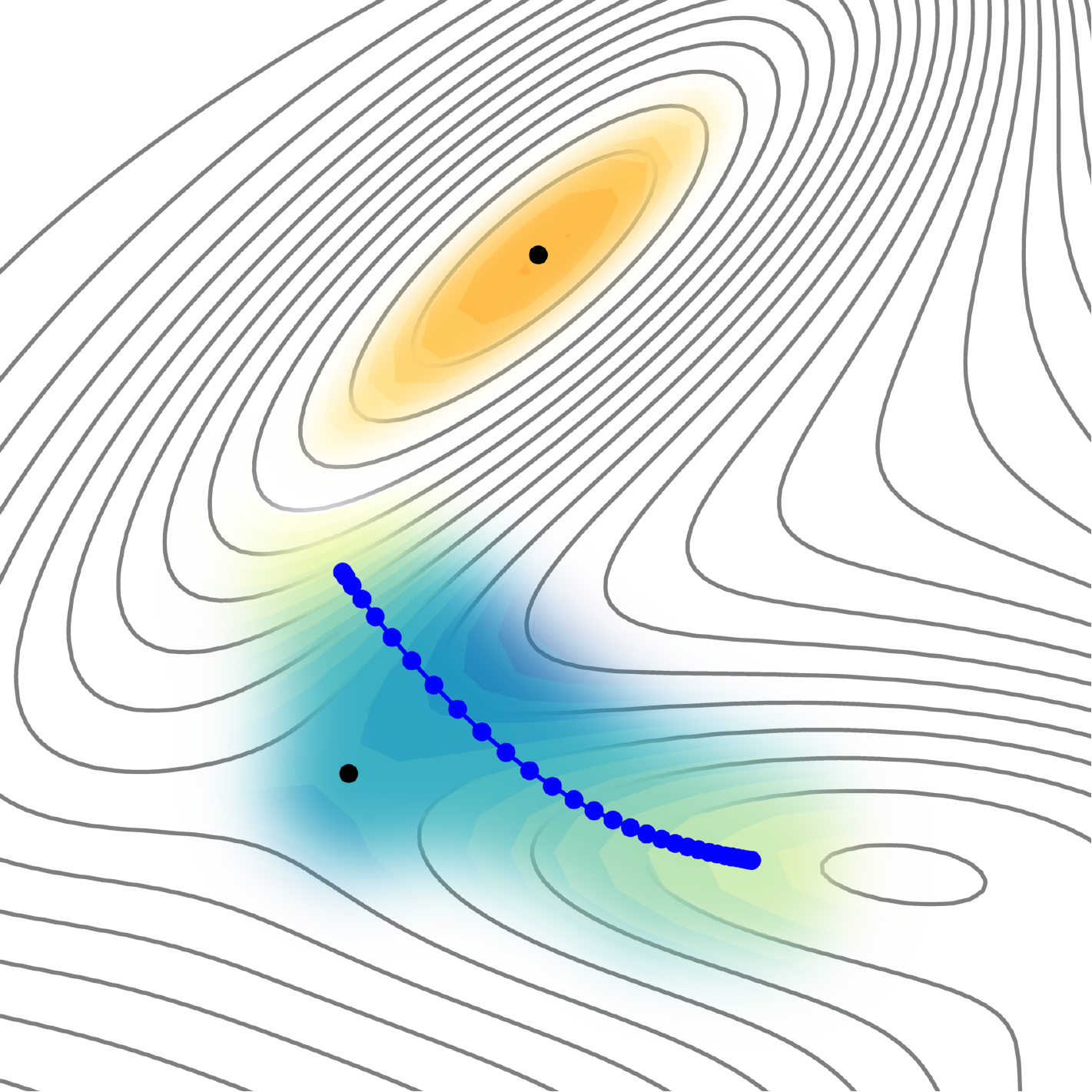}
	\caption{%
The only areas of the PES which need to be accurately known are the those around the instanton pathway and the reactant minima (in order to obtain their partition functions). %
	In this image, they are represented by the coloured areas, whereas those that are not built into the GPR are unshaded.
	The blue points represent the beads along the instanton path, while the black points represent the reactant and the transition state.
	Note that the tunnelling pathway cuts the corner to explore a space far from the transition state.
}
	\label{fig:local-gpr}
\end{figure}

The reason why these fitting procedures are typically difficult
to carry out in practice
is because a PES is a complex high-dimensional function.
For many applications, including molecular dynamics or wave-function propagation,
it is important to have a \textit{globally-accurate} PES.
In particular, if non-physical minima exist in the PES, the dynamics could be attracted there and give nonsensical results.
Instanton theory has a particular advantage in that it only requires knowledge of a small region of the PES,
located along a line representing the dominant tunnelling pathway.
This implies that it might be possible to fit a \textit{locally-accurate} surface around this small region in an efficient manner,
as represented by figure \ref{fig:local-gpr}.
In this way we ensure that no extrapolation is used, but only interpolation, which is expected to be well behaved.

In this paper, we describe how we use Gaussian process regression (GPR) \cite{gpml}
to fit a local representation of the PES
and thereby obtain the instanton rate using only a small number of \textit{ab initio} calculations.
By converging the rate with respect to the number of electronic-structure calculations, it is possible to obtain the same
results as \textit{ab initio} instanton theory, for a fraction of the cost.
In this way, our GPR approach is almost as efficient as a TST calculation, but has the accuracy of a fully-converged \textit{ab initio} instanton calculation.
We are then able to take advantage of recent developments in high-accuracy electronic-structure methods, \cite{Hattig2010}
which might otherwise be too expensive for an on-the-fly calculation.
A similar combination of GPR and path-optimisation has been used successfully by the group of J\'onsson. \cite{Koistinen2016MEP,Koistinen2017NEB}
A number of new developments are necessary for our implementation,
as instanton theory also requires accurate knowledge of Hessians along the path,
and because we apply the approach to gas-phase reactions, we must account for rotational invariance.

In the following, we describe the background theory as well as the particulars of our implementation of the approach.
Results are then presented for two applications and the convergence properties discussed.

\section{Theory}

The results in this paper are computed by combining together a number of different approaches.
Ring-polymer instanton theory is used to evaluate the rate
based on a GPR fit to the PES, which has a training set comprised of coupled-cluster electronic-structure calculations.
It will be necessary to transform some data between different coordinate systems to use an appropriate set for each part of the calculation.
The instanton equations are defined with Cartesians, as are the inputs and outputs of the electronic-structure calculations,
but the GPR is best built using internal coordinates to ensure that it is rotationally invariant.
In this way we formally make no further approximations to instanton theory
and also avoid having to construct a kinetic-energy operator in curvilinear coordinates.

\subsection{Ring-Polymer Instanton Theory}
\label{subsec:rpinst}

In the ring-polymer version of instanton theory, \cite{InstReview}
the dominant tunnelling pathway is represented by a path discretised into $N$ segments.
The points where the segments begin and end are given by Cartesian coordinates, $\mathbf{x}_i$,
called ``beads''.
Because the instanton pathway folds back on itself, only one half of the path need be specified.
\cite{Andersson2009Hmethane,Rommel2011locating}
A path defined by a set of $N/2$ beads, $\{\mathbf{x}_1,\dots,\mathbf{x}_{N/2}\}$, has the associated half-ring-polymer potential
\begin{multline}
\label{rp-potential}
U_{N/2}(\mathbf{x}_1,\dots,\mathbf{x}_{N/2}) \\ = \sum_{i = 1}^{N/2-1} \sum_{j = 1}^{3n} \frac{m_j}{2\beta_N^2\hbar^2}(x_{i, j}-x_{(i-1), j})^2 + \sum_{i = 1}^{N/2}V(\mathbf{x}_i),
\end{multline}
where $x_{i,j}$ is the Cartesian coordinate of the $i$th bead in the $j$th nuclear degree of freedom with associated mass $m_j$.
The number of degrees of freedom is $3n$, where $n$ is the number of atoms.
The spring constants are defined by the temperature, $T$, such that $\beta_N=\beta/N$ and $\beta=(k_B T)^{-1}$.

The instanton configuration is defined as the saddle point of \eqn{rp-potential}
and in practical applications can be located using quasi-Newton geometry optimisers. \cite{Andersson2009Hmethane,RPInst,Rommel2011locating}
These require gradients of the target function
at each iteration, but use update formulae to avoid recomputing the Hessians. \cite{Fletcher}
The gradient of the ring-polymer potential depends on the gradients of the underlying PES at each bead geometry.
In the on-the-fly implementation, these are obtained directly from an electronic-structure package,
but here they are derivatives of the GPR fitted potential.

Once the instanton pathway is optimised, 
the theory accounts for fluctuations up to second order.
Thus in order to evaluate the rate, we require Hessians of each bead.
Again, these can be computed by an electronic-structure package or from the GPR.
The calculation of a Hessian is usually carried out using second-order finite-differences, and is therefore on the order of $3n$-times more expensive than a gradient calculation.

Under the instanton approximation, the rate is given by
\begin{equation}
\label{eq: inst-rate}
k_\mathrm{inst}Q_r = \frac{1}{2\pi\beta\hbar}Q_\mathrm{trans}Q_\mathrm{rot}Q_\mathrm{vib}\exp(-S/\hbar),
\end{equation}
where the action is $S = 2\beta_N\hbar U_{N/2}$ and
explicit expressions for the instanton vibrational, rotational and translational partition functions are given in \Ref{InstReview}.
The result should be converged with respect to the number of beads, $N$.
Typically on the order of $N=100$ beads are used to obtain a rate converged to two significant figures.

\subsection{CCSD(T)-F12 Theory}
\label{sec:CCSD}

For electronic structures where the independent particle model is qualitatively correct,
electronic energies computed at the basis set limit CCSD(T) level of theory are expected
to be accurate to better than 1 kcal/mol for reaction barriers, 0.1~pm for structures and 
5~cm$^{-1}$ for harmonic vibrational wavenumbers.\cite{Tew2011} Until relatively recently, the cost associated
with using the large basis sets traditionally required to access the basis set limit has prevented 
this high level of theory from being routinely used in quantum dynamics simulations, which typically 
require many thousands of energy evaluations. With the maturation of modern F12 explicitly 
correlated theory,\cite{Hattig2011} near basis set limit CCSD(T) energies can now be computed using small 
(triple-zeta) orbital basis, at a cost only 15\% larger than a traditional CCSD(T)/TZ calculation,
and quantum dynamics studies can be performed using near basis set limit CCSD(T) Born-Oppenheimer
potential energy surfaces on a routine basis. 

In F12 theory the standard manifold of correlating orbitals $\vert ab\rangle$ that parameterise 
two body correlation functions in pair theories is supplemented with one geminal basis function 
per occupied orbital pair $ij$, chosen to directly model the coulomb hole in the first-order pair 
correlation function
\begin{align}
\vert \mu_{ij} \rangle = \sum_{a<b} t_{ij}^{ab} \vert ab \rangle + \sum_{k<l} c_{ij}^{kl} \hat Q f(r_{12}) \vert kl \rangle.
\end{align}
The correlation factor $f(r_{12})$ is chosen to be a linear combination of Gaussians\cite{Tew2005} fit to an exponential function\cite{Tenno2004}
with a length-scale of 1~$a_0$, appropriate for valence electrons, and the many electron integrals that arise
due to the explicit dependence of on the interelectronic distance $r_{12}$ and the presence of the strong orthogonality
projector $\hat Q$ are decomposed into one and two-electron components by inserting approximate resolutions of the identity.\cite{Valeev2004}
The coefficients $t_{ij}^{ab}$ are optimised in the presence of fixed geminal contributions,\cite{Tew2008} to reduce geminal basis set
superposition error,\cite{Tew2006} with coefficients chosen to satisfy the first-order singlet and triplet cusp conditions.\cite{Bokhan2009}
Small but numerically expensive geminal contributions to the energy Lagrangian function are neglected if they rank higher than 
third order in perturbation theory,\cite{Kohn2010} resulting in the CCSD(T)(F12*) approximation.\cite{Hattig2010} 
In this work we use the Molpro electronic structure package\cite{Molpro}
and are restricted to using the slightly less accurate CCSD(T)-F12b approximation\cite{Adler2007} where geminal contributions
from third order ring diagrams are also neglected. Nevertheless, CCSD(T)-F12b energies computed in a TZ basis set are within
within 0.2 kJ/mol per valence electron of the CCSD(T) basis set limit and retain the intrinsic accuracy of the wavefunction ansatz.\cite{Tew2016}

\subsection{Gaussian Process Regression (GPR)}
\label{subsec:gpr}
Gaussian process regression is a machine learning algorithm which can be used to efficiently generate complex hypersurfaces with limited data.\cite{gpml}
Recent work has applied this technique for constructing potential-energy surfaces \cite{kolb2017, cui2016, Alborzpour2016} and determining minimum energy paths \cite{Koistinen2016MEP,Koistinen2017NEB} at a much lower computational cost.
In this paper, a local representation of the PES is constructed around the
instanton pathway and used to evaluate reaction rate constants.

Before carrying out the construction of a local PES with GPR,
we first note that we have to utilise an internal coordinate system that accounts for rotational invariance.
We define this internal coordinate system as $\mathbf{q}=\mathbf{q}(\mathbf{x})$, where $\mathbf{x}$ is a set of Cartesian coordinates.
This transformation to a rotationally invariant coordinate system is defined in section \ref{subsec:internals}.

In the simplest case, the training set consists of
known values of the potential, $V(\mathbf{q}_j)$, at the $M$ reference points $\{\mathbf{q}_1, \dots, \mathbf{q}_M$\}. 
This defines the column vector $\mathbf{y}$ with elements
\begin{equation}
\label{eq: potentials}
y_j = \epsilon + V(\mathbf{q}_j),
\end{equation}
where $\epsilon$ is an energy shift chosen such that the average of these elements is approximately zero.
Noting that the derivative of a Gaussian process is also a Gaussian process, \cite{gpml, Koistinen2016MEP,Koistinen2017NEB}
it is also possible to include gradients and Hessians into the training set as described in \Ref{Koistinen2016MEP}.

The potential for an unknown point $\mathbf{q}^*$ can be predicted from GPR as
\begin{equation}
\label{eq: gpr}
V(\mathbf{q}^*) = -\epsilon + \sum_{j = 1}^{M} k(\mathbf{q}^*, \mathbf{q}_j) \: w_j,
\end{equation}
where $k(\mathbf{q}_i, \mathbf{q}_j)$ is a covariance function for the prior.
We chose a squared-exponential covariance function with length-scale $\gamma$ and a prefactor $f$:
\begin{equation}
\label{eq: cov-matrix}
k(\mathbf{q}_i, \mathbf{q}_j) = f^2 \exp(-\frac{1}{2\gamma^2}|\mathbf{q}_i - \mathbf{q}_j|^2).
\end{equation}
The elements, $w_j$, of the vector, $\mathbf{w}$,
are determined by solving the linear equations
\begin{align}
\label{eq: solve-weights}
(\mathbf{K} + \sigma^2\mathbf{I}) \: \mathbf{w} &= \mathbf{y},
\end{align}
where the covariance matrix is defined by $K_{ij} = k(\mathbf{q}_i, \mathbf{q}_j)$.
By differentiating \eqn{eq: gpr}, one obtains expressions for the gradient and Hessian of the PES.
Because the covariance function is smooth, these derivatives are always well defined.

$\sigma$ is a noise term, which is introduced to avoid overfitting,
and should be chosen to be the expected self-consistent error in the reference data.
Together $f$, $\sigma$ and $\gamma$ are known as hyperparameters.
Their values can be optimised by maximising the log marginal likelihood,
\begin{equation}
	\label{eq:log-like}
	\Theta = -\frac{1}{2}\mathbf{y}^T \mathbf{w} - \frac{1}{2}\log{|\mathbf{K} + \sigma^2\mathbf{I}|} - \frac{M}{2}\log{2\pi}.
\end{equation}
Alternatively, one can also optimise the hyperparameters through the minimisation of errors by cross-validation. \cite{gpml}

The method above allows us to construct a local PES from a training set of $M$ points.
In our implementation, the general idea is as follows.
Firstly, we construct an approximate PES with GPR using a small number of points, and then optimise the ring polymer based on this PES.
After this, we refine the PES by adding new \emph{ab initio} evaluations of points along the previously predicted ring-polymer configuration.
Using the refined PES, we obtain a new ring-polymer configuration and then compare it with the previous one to check if the pathway has converged to the true instanton pathway.
If this is not satisfied, the PES is refined again through the addition of \emph{ab initio} evaluations; this is continued iteratively until the convergence is achieved.
The abovementioned scheme is further elaborated in section \ref{sec:algorithm}. 

The general scheme described above is similar to that done by the group of J\'onsson \cite{Koistinen2016MEP,Koistinen2017NEB}, wherein they obtain the minimum energy path using a GPR-aided nudged elastic band (NEB) method.
This appears to have been highly successful, effectively reducing the number of \emph{ab initio} evaluations required by an order of magnitude in comparison to a conventional NEB calculation.
In this paper, we intend to emulate this drastic reduction in computational effort
for locating the instanton pathway and evaluating rates.
As mentioned before, there are some differences in our implementation, such as the need for rotational invariance and accurate knowledge of the Hessians.
We have found that the accuracy of the Hessians returned by GPR is significantly improved by explicitly providing Hessian data into the training set.

\subsection{Non-redundant internal coordinate system}
\label{subsec:internals}
We would like to build the GPR representation of the PES using an internal coordinate system
which is rotationally and translationally invariant.
This is necessary as the relative rotational orientation of individual beads along the instanton pathway is not known \textit{a priori}.
However, we will need to be able to convert the information obtained from the GPR-based PES back into a Cartesian coordinate system
in order to evaluate the instanton rate.
Also note that the data available from electronic-structure packages are in Cartesian coordinates,
which will need to be converted into internal coordinates in order to build the GPR-based PES.

Much recent work into machine-learning algorithms for describing intermolecular forces
has further required permutational invariance.
\cite{Braams2009PES, Csanyi2013, vonLilienfield2012, vonLilienfield2018}
Such advanced approaches could also be applied to our problem.
However, as we only need to fit the potential locally,
it is an unnecessary complication and thus we choose here to neglect permutational symmetry.
For our studies here, this is no inconvenience as we only need compute the instanton rate for one of the equivalent reaction pathways
and multiply the rate by the degeneracy.

A simple translationally and rotationally-invariant coordinate system
for representing molecular geometries
is provided by the $n\times n$ distance matrix \cite{dist-matrix}, defined as
 \begin{equation}
 	\label{eqn:dist_mat}
 	{D}_{ij} = \begin{cases}
 	||\vec{r}_{i} - \vec{r}_{j}||^{-1}, \quad &i > j\\
 	0,   &i \leq{j}, 
 	\end{cases}
 \end{equation}
 where $\vec{r}_i$ are the three-dimensional Cartesian coordinates of atom $i$,
 such that $\vec{r}_1 = (x_1,x_2,x_3)$, $\vec{r}_2=(x_4,x_5,x_6)$, etc.

Although it is possible to convert data from a Cartesian coordinate system into this set, \cite{Pulay1992internal}
the back transformation is not well defined,
as the internal coordinates are redundant.
In order to obtain a non-redundant set of internal coordinates,
we follow the approach of 
Baker et al. \cite{delocalised-internals}
Firstly, we unravel the matrix $\mathbf{D}$ to give the coordinates as a vector of length $n^2$,
\begin{equation}
	\label{eqn:vectorise}
	\mathbf{d} = \begin{bmatrix}
	D_{11} & D_{12} & \dots & D_{21} & D_{22} & \dots & D_{nn}
	\end{bmatrix}^T.
\end{equation}
The $\mathbf{B}$ matrix is defined to describe how changes in the Cartesian coordinates affects these redundant coordinates
as %
\begin{align}
	\mathbf{B} &= \frac{\partial \mathbf{d}}{\partial \mathbf{x}}
			   = \begin{bmatrix}
			   \frac{\partial D_{11}}{\partial r_{11}} & \dots  &  \dots & \dots & \frac{\partial D_{11}}{\partial r_{n3}} \\
			   \vdots & \vdots & \vdots & \vdots &\vdots \\
			   \frac{\partial D_{nn}}{\partial r_{11}} & \dots & \dots & \dots & \frac{\partial D_{nn}}{\partial r_{n3}}
			   \end{bmatrix}. %
\end{align}
The elements of this $n^2\times 3n$ matrix are given explicitly by
\begin{equation}
\label{eqn:bmat3}
\frac{\partial {D}_{ij}}{\partial {r}_{k\alpha}} = \begin{cases}
-({r}_{i\alpha} - {r}_{j\alpha}) ||\vec{r}_{i} - \vec{r}_{j}||^{-3}, \quad & k = i > j \\
({r}_{i\alpha} - {r}_{j\alpha}) ||\vec{r}_{i} - \vec{r}_{j}||^{-3}, \quad & i > j = k \\
0, \quad & \text{otherwise}
\end{cases}
\end{equation}
where $\alpha$ runs over the indices of three-dimensional space.

A square matrix, $\mathbf{G} = \mathbf{BB}^T$, is formed and then diagonalised
to obtain the eigenvalues and eigenvectors.
The non-redundant eigenvectors are those corresponding to the nonzero eigenvalues (of which there will be $3n-6$ for a nonlinear isolated molecule),
whereas the redundant eigenvectors have zero eigenvalues. %
The non-redundant eigenvectors are collected into the columns of a matrix, $\mathbf{U}$.
With this, we can now transform $\mathbf{d}$ into a non-redundant coordinate system, 
defined by
\begin{equation}
	\label{eqn:dist-transform}
	\mathbf{q} = \mathbf{U}^T\mathbf{d}.
\end{equation}
It is this internal coordinate system which is used to build the GPR representation.

Note that the matrix $\mathbf{U}$ is built only once at a reference geometry
and used to define the transformation to $\mathbf{q}$ at all other geometries.
The reference geometry used in our studies was the transition-state, although this is not a requirement.
The same $\mathbf{U}$ matrix is then used for new geometries to give a consistent definition of the internal coordinates $\mathbf{q}=\mathbf{q}(\mathbf{x})$.

Therefore the required relationship between the internal coordinates and Cartesians is given by $\rmd \mathbf{q} = \mathbf{B}_\mathbf{q} \rmd \mathbf{x}$,
where
\begin{equation}
	\label{eqn:newB}
	\mathbf{B}_{\mathbf{q}} = \pder{\mathbf{q}}{\mathbf{x}} = \mathbf{U}^T \mathbf{B}.
\end{equation}

The gradient and Hessian in the non-redundant internal coordinate system are defined as 
\begin{align}
	\label{eq:int_gradhess}
	\mathbf{g}_\mathbf{q} &= \pder{V}{\mathbf{q}} &
	\mathbf{H}_\mathbf{q} &= \pders{V}{\mathbf{q}}{\mathbf{q}},
\end{align}
and similarly, in Cartesian coordinates,
\begin{align}
	\label{eq:cart_gradhess}
	\mathbf{g}_\mathbf{x} &= \pder{V}{\mathbf{x}} & 
	\mathbf{H}_\mathbf{x} &= \pders{V}{\mathbf{x}}{\mathbf{x}}.
\end{align}

Given a geometry $\mathbf{x}$ to define the appropriate orientation,
gradients and Hessians obtained from the GPR in internal coordinates
can be transformed back to Cartesian coordinates.
Obtained using the chain rule, the transformations are defined by
\begin{align}
	\label{eqn:cart-transform}
	&\mathbf{g}_{\mathbf{x}} = \mathbf{B}_\mathbf{q}^T \: \mathbf{g}_{\mathbf{q}} \\
	&\mathbf{H}_{\mathbf{x}} = \mathbf{B}_\mathbf{q}^T \:\mathbf{H}_{\mathbf{q}} \:\mathbf{B}_\mathbf{q} + \mathbf{g}_{\mathbf{q}}^T\frac{\partial \mathbf{B}_\mathbf{q}}{\partial \mathbf{x}},
\end{align}
where $\pder{\mathbf{B_q}}{\mathbf{x}}=\mathbf{U}^T\pder{\mathbf{B}}{\mathbf{x}}$.

In order to transform the 
gradients and Hessians obtained from electronic-structure calculations
into the $\mathbf{q}$ coordinate system,
these equations need to be inverted.
However, as $\mathbf{B_q}$ is not a square matrix, we need to define the generalised inverse as
\begin{equation}
	\label{eqn:b-inv}
	(\mathbf{B}_\mathbf{q}^T)^{-1} = (\mathbf{B}_\mathbf{q}\mathbf{B}_\mathbf{q}^T)^{-1}\mathbf{B}_\mathbf{q}.
\end{equation}
The required transformations are
\begin{align}
	\label{eqn:grad-transform}
	& \mathbf{g}_{\mathbf{q}} = (\mathbf{B}_\mathbf{q}^T)^{-1}\mathbf{g}_{\mathbf{x}} \\
	\label{eqn:hess-transform}
	& \mathbf{H}_{\mathbf{q}} = (\mathbf{B}_\mathbf{q}^T)^{-1}\bigg[\mathbf{H}_{\mathbf{x}} - \mathbf{g}_{\mathbf{q}}^T\frac{\partial \mathbf{B}_\mathbf{q}}{\partial \mathbf{x}}\bigg][(\mathbf{B}_\mathbf{q}^T)^{-1}]^T.
\end{align}

These equations define all the necessary transformations needed
for converting the \textit{ab initio} data into reduced coordinates,
and for converting it back to a Cartesian system at a given orientation.

\section{Method}
\label{sec:algorithm}

Our aim is to reproduce the same result to an \textit{ab initio} instanton calculation performed on the fly.
As with these calculations, 
we must therefore consider convergence with respect to $N$.
For our new approach based on GPR, we must also simultaneously converge the result with respect to the number of points in the training set.

Here we outline our standard protocol for computing converged instanton rates using GPR.
This consists of two parts: first, in which the instanton pathway is located, and second,
in which the fluctuation terms are converged to yield the final instanton rate.
We have attempted to design this protocol to be stable and efficient.
In our study, we have found that this protocol posed no significant problems for the systems tested here.
In future studies, one could consider improvements which may increase the efficiency further.
In a realistic working environment, a researcher has the freedom to add information to the GPR however they like until the result is converged.

Our protocol is designed for the case that single-point \textit{ab initio} calculations are by far the most expensive part of the calculation.
We also assume that Hessian calculations are orders of magnitude slower than potential or gradient evaluations.
This is commonly the case for many electronic-structure methods, especially if the Hessians are computed using finite differences.
The efficiency of our protocol should thus be measured in terms of the number of \textit{ab initio} calculations required, 
and in particular the number of Hessians.
We show these figures for specific examples in the next section.

The protocol described below is intended for a calculation of a single instanton rate at a given temperature,
as is the approach used in the \ce{H + CH4} benchmarks we present in \secref{HCH4}.
If, as is common, one needs the rate at multiple temperatures, it is recommended to start just below the crossover temperature, $T_c$.
The optimised instanton can be used as the initial guess and GPR training set for a calculation at a lower temperature.
We use this more efficient approach for our \ce{H + C2H6} calculations in \secref{HC2H6}.

\subsection{Protocol}
\label{subsec:path-conv}

\begin{enumerate}
	\item Optimise the reactants and transition state (using a standard Quantum Chemistry package),
	and obtain gradients and Hessians for the optimised geometries.
		The optimised transition-state geometry in Cartesian coordinates is notated $\mathbf{x}^\ddagger$.
	\item
		By diagonalising the mass-weighted Hessian at the transition-state, calculate the cross-over temperature,
		\begin{equation}
			\label{eqn:crit-temp}
			T_c = \frac{\hbar \omega_b}{2\pi \: k_B},
		\end{equation}
		where $\omega_b$ is the magnitude of the imaginary frequency.

	\item An initial guess for the instanton configuration is obtained using \cite{Rommel2011locating}
	\begin{equation}
	\label{eq: inst-guess}
	\mathbf{x}_i = \mathbf{x}^{\ddagger} + \Delta \cos\bigg(\frac{2\pi i}{N}\bigg) \mathbf{z} \qquad i=\{1,\dots,N/2\},
	\end{equation}
	where $\mathbf{z}$ is the normalised non-mass-weighted eigenvector corresponding to the imaginary mode at the transition state
	and $\Delta$ is a user-defined spread of points.
	Typically we choose $\Delta\sim 0.1$ \AA\ and $N=16$ for an initial guess.
	\label{spread}
	
	However, if previous instanton optimisations at a higher temperature have been performed successfully,
	these configurations usually provide a better initial guess.
	\item Calculate \emph{ab initio} potentials and gradients for the points obtained in step \ref{spread}.
	\item Repeat until convergence:
	\begin{enumerate}
		\item \label{first1}
		Optimise hyperparameters using methods defined previously under section \ref{subsec:gpr}.
		\item Starting with a low number of beads $N$, locate the ring-polymer path by increasing $N$
		until the action $S/\hbar$ is converged to 2 decimal places.
		\item Check if the mean bead displacement $\Delta x = \frac{1}{N}\sum_{i=1}^{N}||\mathbf{x}_{i}^\text{new} - \mathbf{x}_{i}^\text{old}|| < PC$, where $PC$ corresponds to the path convergence limit. Also check that the convergence of the action $|S^\text{new}- S^\text{old}|/\hbar \leq 10^{-2}$.
		\begin{itemize}
			\item If this is satisfied, this means that the ring-polymer path has converged.
			Continue to step~\ref{final}.
			\item Otherwise provide new inputs (\emph{ab initio} energies and gradients) to the GPR training set along the current ring polymer 
			(i.e.\ increase the number of training points $M$) and then go back to step \ref{first1}.
		\end{itemize} 
	\end{enumerate}
\item \label{final}
Repeat until convergence:
	\begin{enumerate}
		\item \label{first2}
		Provide a couple of new points along the converged instanton pathway to the GPR training set, this time also including Hessians.
		\item Optimise hyperparameters using methods defined previously under section \ref{subsec:gpr}.
		\item Locate the instanton pathway and calculate the rate, $k$, increasing $N$ until this converges.
		\item Test if $|k^\text{new} - k^\text{old}| / k^\text{new} \leq RC$, where $RC$ corresponds to the rate convergence limit. 
		\begin{itemize}
			\item If this is satisfied, the iterative algorithm is terminated, and the current value of $k$ is taken as the converged instanton rate.
			\item Otherwise, return to step \ref{first2}.
		\end{itemize}
	\end{enumerate}
\end{enumerate}

In the following calculations,
we built the GPR using energies in hartrees ($E_h$) and Cartesian coordinates in {\aa}ngstr\"oms ($\si{\angstrom}$).
In these units, typical values used for the length scale were $\gamma \sim 0.3-0.4$,
and for the prefactor, $f = 0.09$.
We specified the noise term differently for potentials, gradients and Hessians,
as $\sigma_V \sim 10^{-6}$, $\sigma_G \sim 10^{-4}$ and $\sigma_H \sim 10^{-3}$.
Convergence limits of $PC = 10^{-2}$ \AA \ and $RC = 10^{-2}$ were used.

We have outlined the simplest protocol which has the desired properties of 
converging the instanton rate without needing a large number of \textit{ab initio} calculations.
However, it is not necessarily the optimal choice for all problems.
In particular, 
it should be noted that in this work, new information is provided to the GPR training set at 
the positions of beads chosen by hand.
This was done in a systematic way, wherein during the path convergence step, the beads were chosen such that 
they are evenly distributed along the current ring polymer. %
Once the path is converged, beads where Hessians are to be included were chosen in a similar manner 
(ie. evenly distributed along the converged pathway). 
There may be better ways of providing new information to the GPR training data;
for instance one can evaluate the expected fitting error along the current pathway
and then provide points at the areas with high variance.
By being more selective, one can potentially further reduce the number of \textit{ab initio} calculations required.

\section{Results}
The method described above was applied to the following two systems:
\begin{itemize}
	\item \ce{H + CH4 -> H2 + CH3}
	\item \ce{H + C2H6 -> H2 + C2H5}.
\end{itemize}

The first is a standard benchmark reaction for testing quantum rate theories
and has been studied with various methods including
MCTDH, \cite{Wu2004HCH4,Welsch2012HCH4}
ring-polymer molecular dynamics,\cite{Suleimanov2011HCH4}
quantum instanton, \cite{Karandashev2015QI}
as well as ring-polymer instanton theory. \cite{Andersson2009Hmethane,HCH4,MUSTreview}

The second reaction is beyond the current limits of exact quantum mechanics unless reduced dimensionality models are used.
Using the GPR formalism, we are able to here present a converged \textit{ab initio} instanton rate for the first time.
We compare these results with those predicted by other semiclassical methods.

\subsection{\ce{H + CH4}}
\label{sec:HCH4}
An on-the-fly \textit{ab initio} instanton calculation has been done by one of us for this polyatomic reactive system.\cite{HCH4}
Here, we use this reaction as a benchmark case for our GPR-aided instanton calculation and show that we are able to obtain 
the same result as an on-the-fly calculation with a significant reduction in the number of potentials, gradients, and most importantly, Hessians required.

The electronic-structure method used in \Ref{HCH4} was RCCSD(T)-F12a/cc-pVTZ
and we use exactly the same method for the training set for the GPR.
Note that in this paper, as well as in \Ref{HCH4}, this incomplete basis set is also used to define the energy of the isolated reactant \ce{H} atom.
The results in \Ref{HCH4} were computed using $N=128$, which we also use here.
This required the calculation of 64 \textit{ab initio} potentials and gradients per iteration of the instanton optimisation scheme.
Because approximately 10 iteration steps usually required for an instanton optimisation, 
about 640 gradients were computed in addition to the 64 Hessians once the instanton had been optimised.
To account for the indistinguishability of the \ce{H} atoms, the instanton rate formula is multiplied by 4. %

We followed the protocol outlined in the previous section
independently for three different temperatures.
This allows us to accurately determine the computational effort required for a converged rate.

\begin{table}
	\small
	\caption{Convergence of the instanton path with iteration of protocol step 5.
	The number of potentials (V), gradients (G), and Hessians (H) included in the GPR training set is explicitly noted.
	Here the single Hessian in the training set corresponds to that of the transition state.
	}
	\label{tbl:path-converge}
	\begin{tabular*}{0.48\textwidth}{@{\extracolsep{\fill}}lllll}
		\hline
		T(K)		& Iteration	& Training set	 		 	& $S/\hbar$	& $\Delta x$ ($10^{-3}$ \AA) \\
		\hline
		300			& 1			& 10V, 10G, 1H     	   	& 25.167	&   -		\\
					& 2			& 19V, 19G, 1H 			& 25.243	& $41.7$	\\
					& 3			& 28V, 28G, 1H			& 25.249	& $1.78$	\\ \hline
		250			& 1			& 10V, 10G, 1H			& 28.957	& -			\\ 
					& 2			& 19V, 19G, 1H			& 29.244	& $37.3$	\\ 
					& 3 		& 28V, 28G, 1H			& 29.301	& $3.17$	\\ 
					& 4			& 37V, 37G, 1H			& 29.274	& $1.77$	\\	
					& 5 		& 46V, 46G, 1H			& 29.278	& $0.13$	\\ \hline
		200			& 1			& 10V, 10G, 1H			& 32.586	& - \\
					& 2			& 19V, 19G, 1H			& 33.871	& $70.1$ 	\\
					& 3			& 28V, 28G, 1H		 	& 33.945	& $2.54$	\\ 
					& 4			& 37V, 37G, 1H			& 33.904	& $1.90$	\\ 
					& 5			& 46V, 46G, 1H			& 33.907 	& $0.70$	\\ \hline
	\end{tabular*}
\end{table}

In table \ref{tbl:path-converge}, the rows correspond to iterations of step 5 of the protocol, in which the pathway is optimised by adding more potentials and gradients to the GPR training set.
The action is seen to converge to two decimal places after only a few iterations.
Here, this was done with fewer than 50 potentials and gradients for all three temperatures.
This means that a reduction in the number of gradient evaluations by an order of magnitude has been achieved.

\begin{figure}
	\centering
	\includegraphics[width=\columnwidth, keepaspectratio]{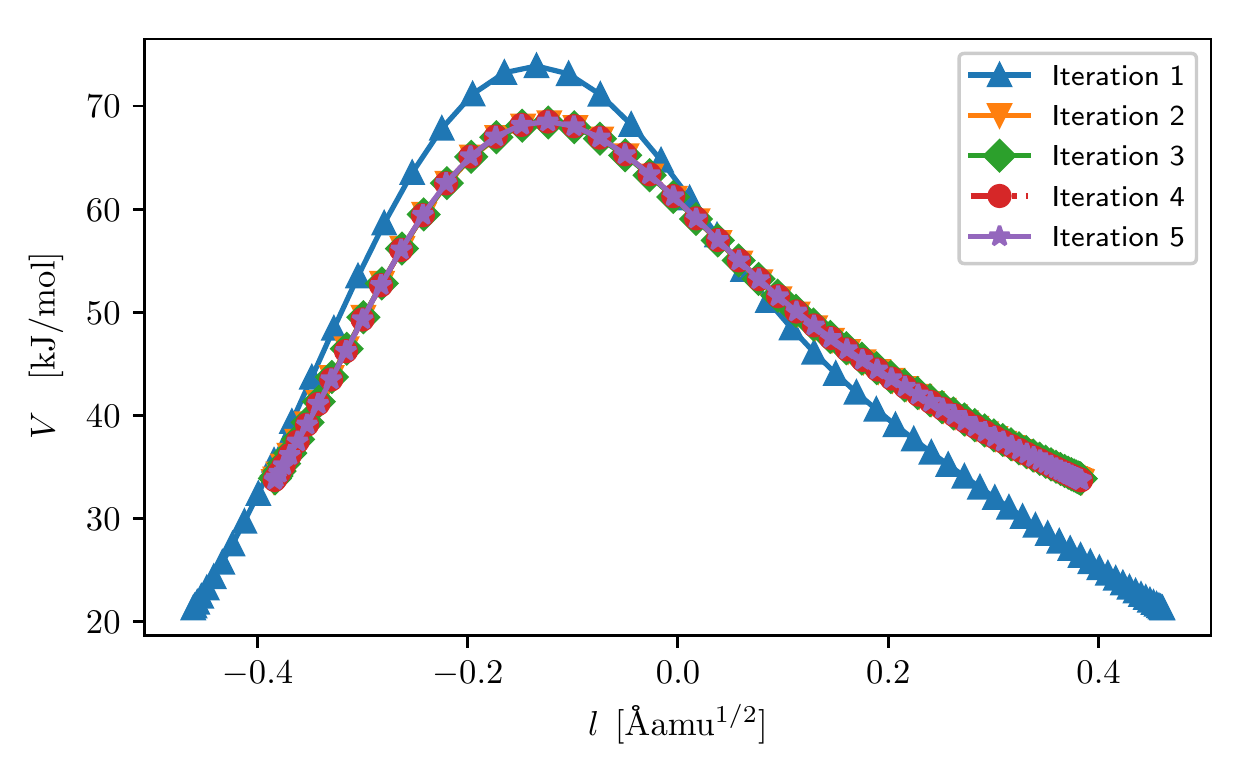}
	\caption{Convergence of the ring-polymer instanton at 200 K for \ce{H + CH4}.
	The initial GPR training set was defined by Eq.~\ref{eq: inst-guess}.
	The ring-polymer beads are plotted as a function of their potential energy and the path length, $l$, as defined by Eq.~\ref{eq:pathlength}.
	}
	\label{fig:inst-path}
\end{figure}

This fast convergence is also represented in figure \ref{fig:inst-path},
where it is seen that, at the lowest temperature studied, the pathway already has the correct shape after the second iteration.
In this figure, the potential along the pathway is plotted as a function of cumulative mass-weighted path length, 
\begin{equation}
	\label{eq:pathlength}
	l_i = \sum_{i'=1}^{i}\sqrt{\sum_{j=1}^{3n}m_j (x_{i'+1, j} - x_{i', j})^2}.
\end{equation}
It should be noted that the plots are shifted such that it is centred around $l = 0$. 

\begin{table}
	\small
	\caption{The rate obtained from the GPR-based instanton calculations are given
	as the information provided to the GPR training set is increased.
	The error is measured relative to the on-the-fly \textit{ab initio} results of \Ref{HCH4}.
	Note that for the rate calculation, one further Hessian is needed at the reactant geometry, but that this is not included in the GPR training set.}
	\label{tbl:trials}
	\begin{tabular*}{0.48\textwidth}{@{\extracolsep{\fill}}llll}
	\hline
	T(K)		& Training set & $k$(cm$^3$ s$^{-1}$)	& Relative Error \\
	\hline
	300			& 28V, 28G, 1H       & 5.49(-19)   	    &   220\%		\\
				& 31V, 31G, 4H		 & 1.72(-19)		&	1.2\%\\ 
				& 33V, 33G, 6H		 & 1.69(-19)		& 	$<$ 1\% \\ 
			&\textbf{\Ref{HCH4}} 	 & \textbf{1.70(-19)}		& -			\\ \hline
	250			& 46V, 46G, 1H		 & 4.25(-20)		& 790\%	\\
				& 49V, 49G, 4H		 & 4.74(-21)		&	-1.3\%	\\
				& 51V, 51G, 6H		 & 4.80(-21)		&	$<$ 1\%	\\
			&\textbf{\Ref{HCH4}}. 	 & \textbf{4.80(-21)}		& -			\\ \hline	
	200			& 46V, 46G, 1H		 & 1.68(-20)		&	$>$ 1000\%	\\
				& 49V, 49G, 4H		 & 0.98(-22)		&	-10\%	\\
				& 51V, 51G, 6H		 & 1.07(-22)		&	-1.8\%  \\
				& 53V, 53G, 8H		& 1.08(-22)			& $<$ 1\% \\
			&\textbf{\Ref{HCH4}}. 	 & \textbf{1.09(-22)}		& -			\\ \hline			
	\end{tabular*}
\end{table}

In table \ref{tbl:trials}, the GPR model is further refined,
as described in step 6 of the protocol,
by providing more observations
(i.e.\ more \textit{ab initio} potentials, gradients and Hessians) to the GPR training set.
Our findings show that it is necessary to include a few Hessians directly into the GPR training set,
and that the transition-state Hessian alone is not sufficient to describe the fluctuation terms.
Note that at low temperatures, the GPR requires slightly more Hessians to converge the rate.
This is due to the fact that the instanton stretches out more at lower temperatures,
thus meaning that GPR needs more information as the instanton covers a larger area of the PES.

The convergence is fast and 
it takes no more than 6 Hessians to converge the rates for all temperatures to less than 2\% of that of the \textit{ab initio} calculation.
This is a remarkable improvement in terms of computational effort required over the \textit{ab initio} instanton calculations
as Hessian calculations account for a huge percentage of the computational effort required.
Having reduced the number of Hessians required from 64 to 6,
the reduction in computational power needed would allow us to investigate problems involving larger molecules and to also use higher-level electronic-structure methods.

\subsection{\ce{H + C2H6}}
\label{sec:HC2H6}
The H abstraction reaction from ethane follows the same mechanism as abstraction from methane.
From a theoretical point of view, it is of interest as the number of degrees of freedom is significantly higher
such that full-dimensional exact quantum methods are not applicable
and approximations must be made.
There are two types of approximations which can be used to make the simulation tractable.
One makes use of semiclassical dynamics, and the other involves reducing the dimensionality of the system.
The instanton method is an example of the former, as are other semiclassical extensions of transition-state theory \cite{cvbmm, Greene20161D} and ring-polymer molecular dynamics. \cite{cvbmm-rpmd}
Reduced-dimensionality models allow quantum scattering theory to be applied \cite{quantum-scatter}
and can also be combined with semiclassical approaches. \cite{rpmd-alkanes, Greene20161D, Greene2016SCTST}
Experimental results are available at 300~K, \cite{Baulch2013, Siva2012}
but unfortunately not at lower temperatures, where the tunnelling effect is more important.
Here, we compare the results of our instanton rate calculations with other theoretical calculations,
and discuss the relative efficiency of the various methods.

\subsubsection{\textit{Ab initio} calculations}

Due to the efficiency of the GPR-aided instanton approach seen in our benchmark tests,
we are able to use high-accuracy and computationally expensive electronic-structure methods.
The method we choose is UCCSD(T)-F12b as discussed in \secref{CCSD}.
Table \ref{table:barrier} shows the predictions for barrier heights, $V^\ddagger$, and imaginary frequencies, $\omega_b$,
with increasingly large basis sets.
Hessians with cc-pVQZ and cc-pV5Z basis sets were not evaluated due to the large amount of computational resources which would be required.
However, we can see that, the cc-pVTZ-F12 reproduces almost the same barrier height as cc-pV5Z,
in accordance with the study by Spackman et al.\cite{f12basis-convergence}
which suggested that cc-pVnZ-F12 basis sets have similar performance to cc-pV(n+2)Z basis sets (where n = D, T, Q, etc.)
in terms of results when using CCSD(T)-F12.
Hence in the following calculations, we will use the cc-pVTZ-F12 basis set. %

\begin{table}
	\small
	\caption{Barrier heights and imaginary frequencies for \ce{H + C2H6} using increasingly larger basis sets at UCCSD(T)-F12b level.}
	\label{table:barrier}
	\begin{tabular*}{0.48\textwidth}{@{\extracolsep{\fill}}lll}
		\hline
		Method   		  			 & $V^\ddagger (\mathrm{kJ \:mol^{-1}})$        & $\omega_b (\mathrm{cm^{-1}})$		\\ \hline
		UCCSD(T)-F12b/cc-pVDZ		 & 54.57							& 1398						\\ %
		UCCSD(T)-F12b/cc-pVTZ		 & 51.24					 		& 1461 		   				\\ 
		UCCSD(T)-F12b/cc-pVTZ-F12	 & 50.03							& 1469						\\ %
		UCCSD(T)-F12b/cc-pVQZ		 & 50.46							& -							\\ 
		UCCSD(T)-F12b/cc-pV5Z		 & 50.07							& -							\\ 
		\hline 
	\end{tabular*}
\end{table} 

\begin{figure}
	\centering
	\includegraphics[width=\columnwidth, keepaspectratio]{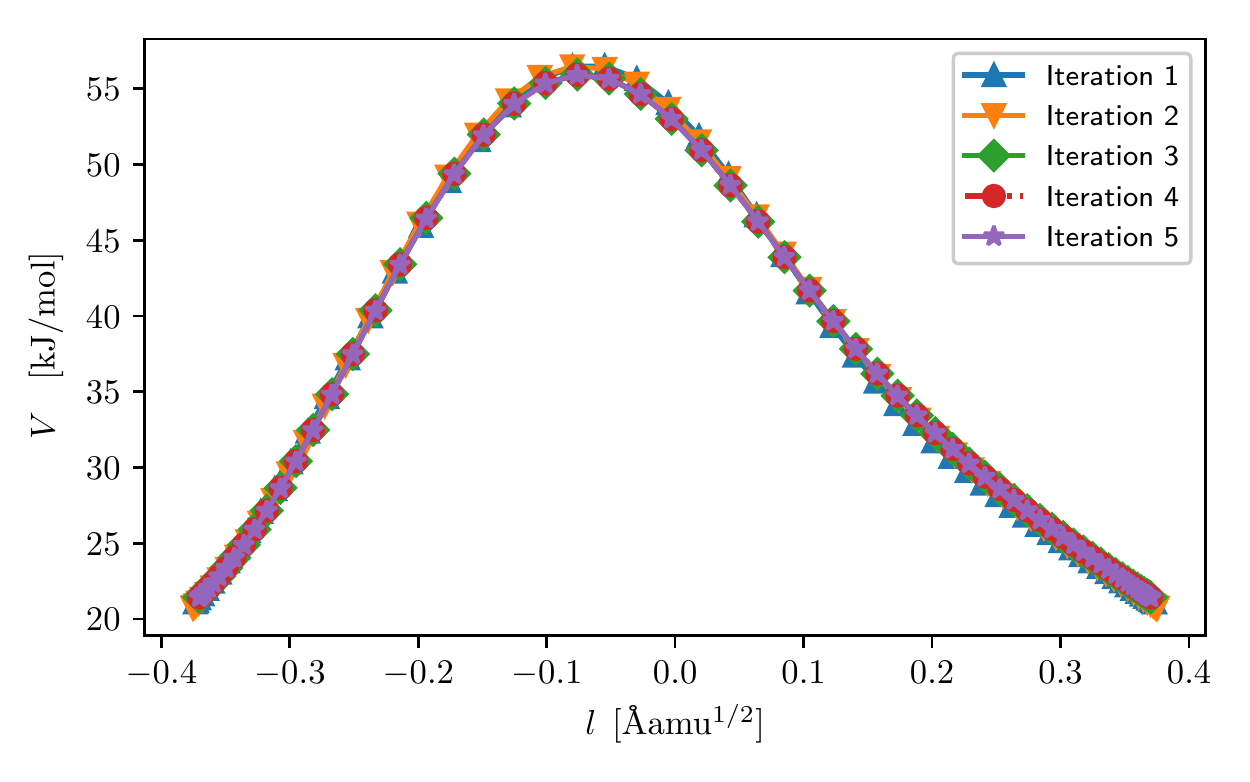}
	\caption{Convergence of ring-polymer instanton at 200 K for \ce{H + C2H6}.
	The initial GPR training set was given by points along the 250 K instanton path.
	The path length, $l$, is defined by equation \ref{eq:pathlength}.
	}
	\label{fig:et-instpath}
\end{figure}

With our chosen method,
the crossover temperature is predicted to be 337 K.
We ran three instanton calculations, first at 300 K, and then used this as a starting point for a calculation at 250 K,
and in the same way for 200 K.
This approach may slightly reduce the number of iterations needed for convergence.
For instance, it can be seen in figure \ref{fig:et-instpath}, that the optimisation of the 200 K instanton
is obtained in only a few iterations and that the path is almost correct even after the first.
The convergence criteria used for this system were similar to that used in the \ce{H + CH4} system.

\begin{figure}
	\centering
	\includegraphics[height=3.5cm, keepaspectratio]{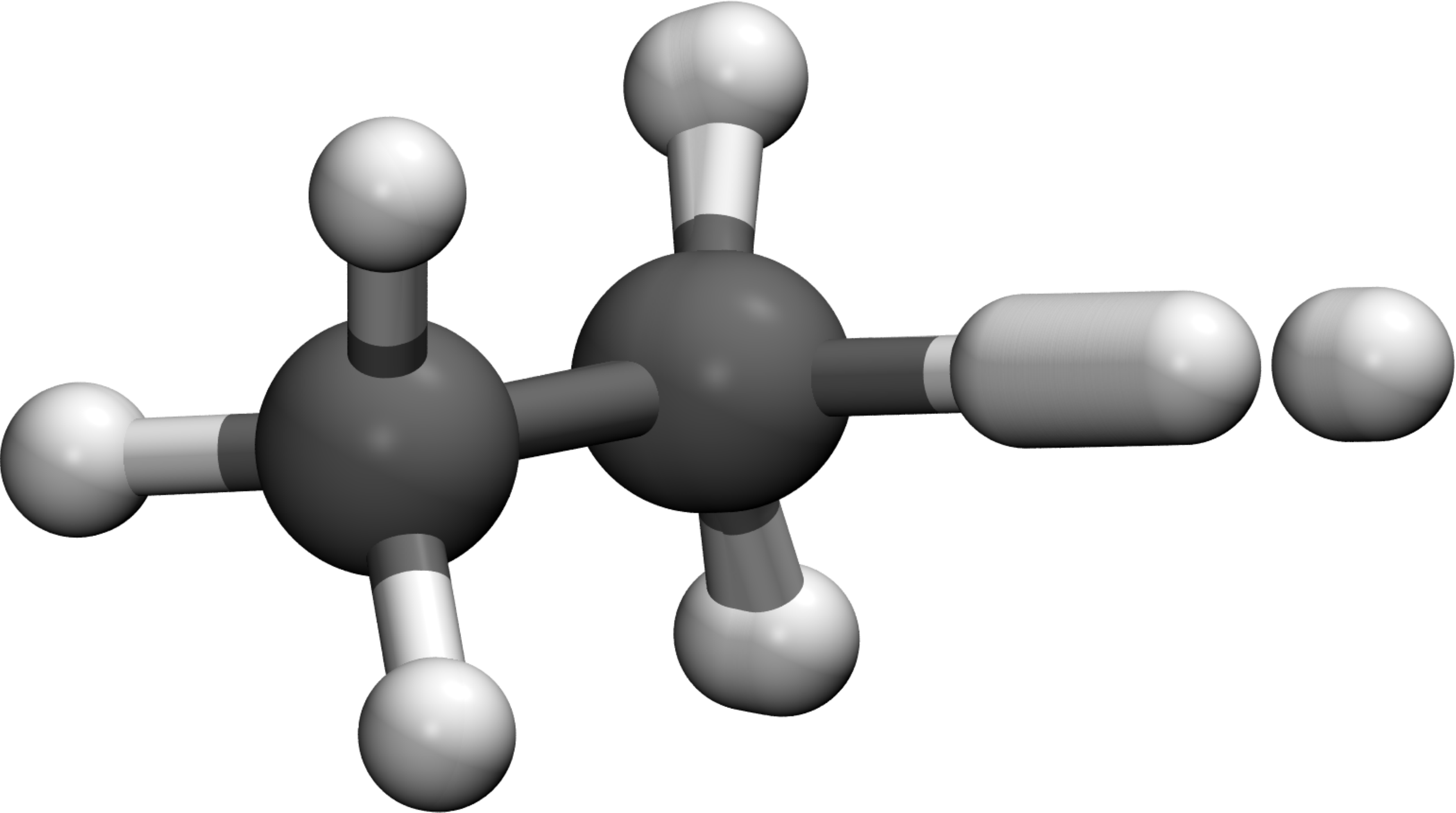}
	\caption{Representation of the ring-polymer instanton for \ce{H + C2H6} at 200~K.}
	\label{fig:instanton}
\end{figure}

The Cartesian representation of the optimised path for the H abstraction from ethane
is shown in figure \ref{fig:instanton}.
It is seen that the mechanism is similar to that of \ce{H + CH4}, shown in \Ref{HCH4},
in that the abstracted hydrogen does most of the tunnelling, and is accompanied by a small movement of its neighbouring hydrogens.
The atoms on the far end of the ethane molecule hardly participate in the instanton at all.
Note, however, that they still make a contribution to the fluctuations, and thus cannot be neglected. \cite{formic}

The results of our GPR-based instanton calculations are presented in table \ref{tbl:ab-init-comparison}.
These rates account for the degeneracy of the reaction by multiplying the formula in \eqn{eq: inst-rate} by a factor of 6. 
The 300 K result was obtained with a training set including 33 potentials and gradients, and 6 Hessians.
Calculations at the lower temperatures of 250 K and 200 K added 
an additional 6 Hessians to the training set
(i.e. at 250 K, training data includes 6 Hessians from 300 K and 6 Hessians from 250 K) 
in order to converge the rates.
This represents a reduction in computational effort by an order of magnitude, similar to what has been observed for \ce{H + CH4}.

\begin{table}
	\caption{Calculated rates (in $\mathrm{cm^3 \: s^{-1}}$) for \ce{H + C2H6} obtained by the GPR-aided instanton method and other direct dynamics methods.
	The tunnelling factor, $\kappa_\text{tun}$, is defined as the ratio between the instanton rate and Eyring TST.}
	\label{tbl:ab-init-comparison}
	
	\begin{tabular*}{0.5\textwidth}{@{\extracolsep{\fill}}lllll}
		\hline
		\multirow{2}{2em}{$T / K$} & \multicolumn{2}{c}{GPR-aided Instanton} & \multirow{2}{5em}{SCTST\textsuperscript{\onlinecite{Greene20161D}}}		& \multirow{2}{5em}{RD-QS\textsuperscript{\onlinecite{quantum-scatter}}}   \\ \cline{2-3} \rule{0pt}{2.5ex}
		 & $\kappa_\text{tun}$ & Rate 		        	\\ \hline 
 		300 			   & 15					& 7.0(-17) 			& 3.88(-17) 						& 6.23(-17)     		 \\
 		250 			   & 38					& 6.4(-18)    		& 9.51(-18)							& 7.97(-18)		 	     \\
 		200 			   & 623				& 5.7(-19)  		& 2.50(-19)  						& 6.69(-19) 		 	 \\    
 		\hline
	\end{tabular*}
	
\end{table}

It is clear from our calculation that tunnelling effect makes a large contribution to the rate, even at 300 K.
This is confirmed by experimental results at this temperature, which in various setups, have been measured to be
$3.13 \times 10^{-17}$ cm$^3$ s$^{-1}$,\cite{Baulch2013} 
or $7.47 \times 10^{-17}$  cm$^3$ s$^{-1}$\cite{Siva2012},
and which both lie in the same order of magnitude as our prediction.
Note that we expect the instanton approach to slightly overpredict the rate (by up to a factor of 2) at 300 K as this lies close to the value of $T_c$. \cite{Faraday}
Unfortunately, no experimental results are available for comparison at lower temperatures
where the tunnelling effect is predicted to increase dramatically.

Table \ref{tbl:ab-init-comparison} also compares our predicted rate
with those of reduced-dimensionality quantum scattering (RD-QS) calculations by Horsten et al. \cite{quantum-scatter} 
and a full-dimensional semiclassical transition state theory (SCTST) rate calculation by Greene et al. \cite{Greene20161D}

The RD-QS calculations utilised a similar electronic structure method to our calculations, albeit with F12a rather than F12b, which gives a barrier height only 0.1 $\mathrm{kJ \: mol^{-1}}$ lower. %
The SCTST calculations employed the CCSD(T)/cc-pVTZ method for the energies at the stationary points, which gives a barrier height 0.4 $\mathrm{kJ \: mol^{-1}}$ lower. 
We expect these differences to lead to only a minor deviation. 

The instanton results are in quite close agreement with RD-QS, where the rates differ by no more than 25\%. 
This is what is typically expected when comparing results obtained with the instanton method and that obtained with exact quantum methods. \cite{MUSTreview}
This confirms that, at least for this system, the reduced-dimensionality approach is not causing an appreciable error in the tunnelling effect.

There is a slightly larger discrepancy between the instanton and SCTST results, \cite{Greene20161D} 
which increases at lower temperatures.
The SCTST rate calculation involved a total of 118 \emph{ab-initio} Hessians at the MP2/cc-pVTZ level,
with energies at stationary points evaluated with CCSD(T)/cc-pVTZ. 
The GPR-Instanton method required only 6 Hessians to converge the rate at each temperature
and thus a high-level of theory for the Hessian calculations can be used as well.
There are two reasons for the discrepancy in the SCTST rates.
One is that lower-level electronic-structure theory was used in \Ref{Greene20161D} for the Hessian calculations.
The second is that at low temperatures, the instanton pathway stretches far from the transition state
and the PES cannot therefore be well represented by a Taylor series around the transition state.

In this case, there are no dramatic differences between the theoretical predictions.
It seems that the \ce{H + C2H6} reaction follows a simple pathway for which reduced-dimensionality models are applicable.
However, we expect that for more complex reactions there will be a larger discrepancy and that,
in many cases, the full-dimensional instanton theory will be the most accurate.

\subsubsection{Results on a fitted PES}

In order to get an idea of the accuracy of the instanton approach for this reaction,
we compare instanton rates with those of other semiclassical approaches based on the fitted, global CVBMM potential-energy surface.
\cite{cvbmm}

This PES was constructed by dividing the system into a reactive part which would be treated with semiempirical valence 
bond theory and a non-reactive part treated with molecular mechanics. 
It was parametrised against density functional theory, of which more details can be found in \Ref{cvbmm}. 
The barrier height obtained with the CVBMM PES is 47.90 kJ mol$^{-1}$
and has a predicted crossover temperature of 352 K. 

Table \ref{tbl:cvbmm-comparison} presents rates of three methods, instanton theory (this work), quantum instanton theory (QI) \cite{cvbmm-qi}
and the small curvature tunnelling correction to canonical variational TST (CVT/SCT). \cite{cvbmm}
The tunnelling factors are seen to be about a factor of 2 larger than those from the \textit{ab initio} method,
mainly due to the fact that the CVBMM barrier is too narrow and thus overpredicts the tunnelling factors.

The CVT/SCT rate is in close agreement with that of instanton theory,
which implies that, at least in this case,
the dominant tunnelling pathway, %
is well approximated by the minimum-energy pathway used by CVT/SCT.
It is expected that, in general for more complex reactions,
the instanton method, which defines the tunnelling pathway in a rigorous manner,
will give a more accurate result.

\begin{table}
	\caption{Rate comparison between methods using the CVBMM PES. All rates are in $\mathrm{cm^3 \: s^{-1}}$.} 
	\label{tbl:cvbmm-comparison}
	\small
	\begin{tabular*}{0.48\textwidth}{@{\extracolsep{\fill}}lllll}
		\hline
		\multirow{2}{2em}{$T / K$} & \multicolumn{2}{c}{Instanton} & \multirow{2}{*}{CVT/SCT\textsuperscript{\onlinecite{cvbmm}}} &\multirow{2}{5em}{QI\textsuperscript{\onlinecite{cvbmm-qi}}} \\ \cline{2-3} \rule{0pt}{2.5ex}
		& $\kappa_\text{tun}$ & Rate \\
		\hline
		300 			   & 24					& 1.25(-16) 		& 1.44(-16) 				& 1.15(-16)     		 \\
		250 			   & 80					& 1.46(-17)    		& -							& 1.40(-17)				 	     \\
		200 			   & 1296				& 1.61(-18)  		& 1.90(-18)  				& 1.16(-18)			 		\\    
		\hline
	\end{tabular*}
\end{table}

Unlike the ring-polymer instanton approach, the QI method does not use a steepest-descent approximation
and thus includes anharmonic vibrational effects in full dimensionality.
In order to do this, it samples over a statistically large number path-integral configurations
and would therefore not be a practical computational method
when combined high-level \textit{ab initio} potentials.
Nonetheless, these anharmonic effects only change the rate by less than 50\% at the lowest temperature studied.
This is in agreement with the findings of \Ref{cvbmm-qi} which showed that, at low temperatures,
a small increase in the rate resulted from making a harmonic approximation to the internal rotation.
This confirms that instanton theory gives a reliable prediction of the order-of-magnitude of the rate.
The real advantage of the instanton approach over this method
is that it can be applied to new reactions
without needing to build a global PES at all.

\section{Conclusions}

We have demonstrated how \textit{ab initio} instanton theory can be made efficient
by using GPR to fit the PES locally around the dominant tunnelling path.
This was demonstrated first using the \ce{H + CH4} reaction as a benchmark,
for which we have shown that the number of electronic-structure calculations can be reduced by an order of magnitude,
while converging the rate to within 1\% of the benchmark result.
We then proceeded to evaluate instanton rates for \ce{H + C2H6},
based on UCCSD(T)-F12b/cc-pVTZ-F12 electronic-structure calculations.
Most importantly, the number of Hessians needed for all these calculations is about 6,
which makes the method more efficient than full-dimensional SCTST calculations
and almost as efficient as a classical TST calculation.

When studying a complex network of reactions,
TST is commonly used to obtain a rate for the many possible reaction steps. \cite{Simm2017network}
By evaluating the crossover temperature for each step,
it can be easily determined whether tunnelling is likely to play a role,
and instanton calculations can be run for these steps only.
As there are typically many more steps for which tunnelling is not important, than those for which it is,
the number of \textit{ab initio} calculations needed for the instanton calculations would be small in comparison to the overall total.
In this way, tunnelling can be rigorously accounted for without significantly increasing the computational effort.

In this work, we suggested a simple protocol which, in our tests, showed no particular problems.
We note, however, that it could still be improved in a number of ways which would further increase the efficiency.
For instance, by using estimates of the GPR fitting error,
we could select new points to be added to the training set in a more systematic way.
These could also be used to estimate the fitting error in the rate constant in a similar way to as has been done for TST calculations. \cite{Proppe2016uncertainty}

Other techniques might allow us to reduce the number of high-level calculations by including low-level \textit{ab initio} information into the GPR training set.
One possibility would be to use this low-level information only for the initial iterations to locate the region of space where the instanton is likely to exist on the high-level surface.
The final iteration could be done using only high-level information to ensure convergence to the correct result.
However, one could also consider combining the high- and low-level information in the training set,
as in the dual-level approach. \cite{Meisner2018dual}
By using a larger value of the noise term for the low-level points,
the GPR would then fit itself accurately to the high-level points,
and use the low-level information as a rough guide for the shape.
Typically the frequency calculations from low-level calculations are a good approximation even if the absolute energies are not,
and so most Hessians could be derived from low-level calculations.
One could imagine systematically converging to the correct result by adding more high-level \textit{ab initio} points
such that the accuracy would not be compromised.

We have shown in this paper that we can converge the rate with respect to the number of ring-polymer beads, $N$,
as well as with respect to the number of points included in the GPR training set.
However, the accuracy of our method is still limited by the computational expense of electronic-structure methods,
which are rarely possible to fully converge.
Methods such as F12 have been very useful for increasing this efficiency \cite{Hattig2011}
as well as linear-scaling methods \cite{Riplinger2013linearscaling}
and the use of graphical processing units. \cite{Ufimtsev2008GPU}
Nonetheless, we can say that we have expanded the range of systems
which can be studied with \textit{ab initio} instanton theory
using high-level electronic structure methods. %

We did not find particularly large differences in rate predictions for the \ce{H + C2H6} reaction
between the instanton approach and other theories.
This is due to the rather simple mechanism exhibited by the H abstraction reaction,
which follows a pathway close to the minimum-energy path,
making the CVT/SCT and reduced-dimensionality models valid.
The advantage of instanton theory is that no \textit{a priori} choice of reduced coordinates, or tunnelling coordinate is made.
This makes the approach applicable also to more complex reactions as well as tunnelling splitting calculations. \cite{tunnel}
In these cases it is expected that the instanton path will deviate more strongly from the minimum-energy path,
and the full-dimensional instanton theory
will be required to obtain an accurate prediction. 
The proof of principle outlined in this work for combining GPR with instanton theory
will then be exploited in future studies of new reactions.

\section*{Conflicts of interest}
There are no conflicts to declare.

\section*{Acknowledgements}
This work has been financially supported by the Swiss National Science Foundation (Project No.\ 175696).

%

%

%

%
%

%
%

%
%
%

%
%
%
%
%
%

\end{document}